\documentclass[english,superscriptaddress]{revtex4}
\usepackage[T1]{fontenc}
\usepackage[latin9]{inputenc}
\usepackage{amsmath}
\usepackage{graphicx}

\makeatletter
\@ifundefined{textcolor}{}
{%
 \definecolor{BLACK}{gray}{0}
 \definecolor{WHITE}{gray}{1}
 \definecolor{RED}{rgb}{1,0,0}
 \definecolor{GREEN}{rgb}{0,1,0}
 \definecolor{BLUE}{rgb}{0,0,1}
 \definecolor{CYAN}{cmyk}{1,0,0,0}
 \definecolor{MAGENTA}{cmyk}{0,1,0,0}
 \definecolor{YELLOW}{cmyk}{0,0,1,0}
}

\makeatother

\usepackage{babel}
\begin{document}

\title{Unifying averaged dynamics of the Fokker-Planck equation for Paul
traps}

\author{Arindam Bhattacharjee}
\email{arindam1001@gmail.com}

\affiliation{Department of Mechanical Engineering, Indian Institute of Technology
(IIT), Kanpur - 208016, Uttar Pradesh, India.}

\author{Kushal Shah}
\email{kushals@iiserb.ac.in}

\affiliation{Department of Electrical Engineering and Computer Science, Indian
Institute of Science Education and Research (IISER), Bhopal - 462066,
Madhya Pradesh, India.}

\author{Anindya Chatterjee}
\email{anindya@iitk.ac.in}

\affiliation{Department of Mechanical Engineering, Indian Institute of Technology
(IIT), Kanpur - 208016, Uttar Pradesh, India.}
\begin{abstract}
Collective dynamics of a collisional plasma in a Paul trap is governed
by the Fokker-Planck equation, which is usually assumed to lead to
a unique asymptotic time-periodic solution irrespective of the initial
plasma distribution. This uniqueness is, however, hard to prove in
general due to analytical difficulties. For the case of small damping
and diffusion coefficients, we apply averaging theory to a special
solution to this problem, and show that the averaged dynamics can
be represented by a remarkably simple 2D phase portrait, which is
independent of the applied rf field amplitude. In particular, in the
2D phase portrait, we have two regions of initial conditions. From
one region, all solutions are unbounded. From the other region, all
solutions go to a stable fixed point, which represents a unique time-periodic
solution of the plasma distribution function, and the boundary between
these two is a parabola.
\end{abstract}
\maketitle

\section{Introduction}

Periodically driven systems provide a widely used framework for the
study of many phenomena in plasma physics \cite{Turner,Shah2008,Paul}
and statistical mechanics \cite{Corte,Golding,Shah2017,Rahav}. Some
of these systems could be externally driven by electromagnetic fields
or other time-periodic forces \cite{Shah2008,Turner,Gelfreich} and
others could have some internal parts attached to a spring \cite{Shah2017,Neishtadt}.
Systems which are externally driven do not obey conservation of energy
and could thus experience an unbounded growth of energy \cite{Gelfreich},
which has also been experimentally observed to some extent in a quadrupole
ion traps, also called Paul traps \cite{Ryjkov,Monroe}. A Paul trap
is a simple device used to trap charged particles through use of time-periodic
electric forces \cite{Paul,Saxena,Abraham}. Earnshaw's theorem says
that an electrostatic potential cannot have a local minimum/maximum
in 3D space and hence, cannot be used to confine charged particles
on its own. One way to confine charged particles is to use time-periodic
electric fields as is done in Paul traps \cite{Paul}, which can be
shown to form a local extremum in effective trapping potential in
3D space by use of averaging theory for small values of the trapping
parameters \cite{Paul,Shah2008}. As compared to other particle traps,
a Paul trap is easier to construct and miniaturize, but the collective
charged particle dynamics in a Paul trap is quite complex, particularly
with regard to the phenomenon of rf heating. Essentially, what is
observed in Paul traps is that the ion temperature is usually much
higher than that of the background gas and it is this phenomenon which
is known as rf heating. This is believed to be caused due to the external
periodic driving, but a complete and convincing explanation of this
effect is not yet available. 

In a Paul trap, the externally applied potential is given by \cite{Paul}
\begin{equation}
\Phi\left(x,y,z,t\right)=\frac{U_{0}+V_{0}\cos\left(\omega t\right)}{r_{0}^{2}+2z_{0}^{2}}\left(x^{2}+y^{2}-2z^{2}\right)\label{eq:Paul-Trap-Potential}
\end{equation}
where $U_{0}$ is the DC potential, $V_{0}$ is the AC potential with
frequency $\omega$ and, $r_{0}$ and $z_{0}$ are the trap dimensions.
The expression for $\Phi$ given above satisfies the Laplace equation
in free space, $\nabla^{2}\Phi=0$. Interestingly, the corresponding
electric field components in the three directions $\left(x,y,z\right)$
get decoupled from each other, and hence we could, in principle, analyze
particle dynamics in each direction and then combine them to form
a complete 3D picture. Of course, there are small nonlinear effects
in practical traps which require taking into account interaction between
particle motion in different directions, but they can be ignored in
the first approximation.

The single particle equation in the $x$-direction for a Paul trap
can be written as 
\begin{equation}
\ddot{x}=-f\left(t\right)x,\label{eq:Single-Particle-Eqn}
\end{equation}
where 
\begin{equation}
f\left(t\right)=p-q\cos\left(\omega t\right),\label{eq:E-Force}
\end{equation}
which is the well known Mathieu equation \cite{McLachlan}, and the
parameters $p,q$ are obtained from Eq. \eqref{eq:Paul-Trap-Potential}.
Collective dynamics of charged particles can be modeled by the Vlasov
(or Liouville) equation in the absence of collisions, which holds
true for a very dilute plasma \cite{Nicholson}
\begin{equation}
\frac{\partial P}{\partial t}+v\frac{\partial P}{\partial x}+\frac{eE\left(x,t\right)}{m}\frac{\partial P}{\partial v}=0\label{eq:VlasovEquation}
\end{equation}
where $P\left(x,v,t\right)$ is the plasma distribution function,
$e$ is the ion charge, $m$ is the ion mass and $E\left(x,t\right)$
is the electric field. Ideally, the Vlasov equation should be solved
along with the Maxwell's equations to take care of self-consistency.
However, if the externally applied field is much stronger than the
induced field, the latter can be ignored \cite{Shah2009}. The Vlasov
equation has been solved earlier for the case of a spatially linear
time-periodic electric field and it has been found that the solutions
are typically aperiodic in time, but can also be periodic for certain
specific initial conditions \cite{Shah2008,Banerjee}. This leaves
open the question as to which of these infinitely many distributions
obtained by solving the Vlasov equation actually gets manifested in
an experimental setting. This question cannot be answered from the
Vlasov perspective and one has to consider the effects of inter-particle
collisions.

Inter-particle interactions of the collisional system are usually
modeled using the Fokker-Planck equation given by \cite{Risken,Shah2010,Dutta,Chandra},
\begin{equation}
\frac{\partial P}{\partial t}+v\frac{\partial P}{\partial x}+\frac{eE\left(x,t\right)}{m}\frac{\partial P}{\partial v}=\gamma\frac{\partial vP}{\partial v}+D\frac{\partial^{2}P}{\partial v^{2}}\label{eq:FP-Paul}
\end{equation}
where $\gamma>0$ is the damping coefficient and $D>0$ is the diffusion
coefficient. For the case of a Paul trap, $eE\big/m=-f\left(t\right)x$.
The above equation reduces to the Vlasov equation when $\gamma=0=D$.
Certain solutions of this Fokker-Planck equation for spatially non-uniform
time-periodic electric fields have earlier been found to asymptotically
become time-periodic \cite{Shah2010,Dutta}. But what is currently
not known is whether there are multiple time-periodic solutions that
the system can reach asymptotically, or is there only one unique time-periodic
solution. This question is hard to answer in general due to analytical
difficulties, but when the damping and diffusion coefficients are
small, averaging theory can be applied in principle. In this paper,
we analytically solve the Fokker-Planck equation using averaging techniques
for the case of the spatially linear time-periodic electric field
of a Paul trap and show that for a set of initial conditions, the
plasma distribution reaches a unique time-periodic solution asymptotically.
The averaged equation also admits other (unstable) solutions and behaviors,
which we will present below.

In Sec. \ref{sec:Form-of-the-Solution}, we present the particular
form of the solution of the Fokker-Planck equation that we consider
in this work and Sec. \ref{sec:Solution-of-the-Collisionless} contains
the corresponding solutions of the collision-less case. In Sec. \ref{sec:Averaging-of-the-Collisional},
we present the averaging method used for solving the Fokker-Planck
equation for a collisional plasma and the resulting slow-flow dynamics
is described in Sec. \ref{sec:Dynamics-of-the-Slow-Flow}. Finally,
we end the paper with discussion and conclusion in Sec. \ref{sec:Discussion-and-conclusion}.

\section{Form of the solution\label{sec:Form-of-the-Solution}}

For the electric field expression of a Paul trap, as was shown in
\cite{Shah2010}, Eq. \eqref{eq:FP-Paul} admits solutions of the
form
\begin{align}
P\left(x,v,t\right) & =g\left(t\right)\exp\left[-A\left(t\right)v^{2}-B\left(t\right)xv-C\left(t\right)x^{2}\right]\label{eq:FP-Sol-Form1}
\end{align}
where $g\left(t\right)$, $A\left(t\right)$, $B\left(t\right)$ and
$C\left(t\right)$ are unknown functions of time, to be determined.
Substituting Eq. \eqref{eq:FP-Sol-Form1} in Eq. \eqref{eq:FP-Paul},
and equating coefficients of various powers of $x$ and $v$, we obtain
\begin{align}
\dot{A} & =2\epsilon A-B-4\epsilon A^{2}\nonumber \\
\dot{B} & =\epsilon B-2C+2f\left(t\right)A-4\epsilon AB\nonumber \\
\dot{C} & =f\left(t\right)B-\epsilon B^{2}\label{eq:ABC-Scaled-Eqn}
\end{align}
where the dot represents a derivative with respect to time, the functions
$A,B,C$ have been scaled by a factor of $D\big/\gamma$ in order
to eliminate $D$ from the equation, and $\gamma$ is henceforth denoted
as $\epsilon$, with the understanding that $0<\epsilon\ll1$. In
order to maintain normalization of the plasma distribution $P\left(x,v,t\right)$
for all time, we must have $g=\sqrt{4AC-B^{2}}$. In this paper, we
take $p=0$ and $q>0$. It is important to note here that any solutions
of Eq. \eqref{eq:ABC-Scaled-Eqn}, in principle, provide a solution
to the Fokker-Planck equation, Eq. \eqref{eq:FP-Paul}. It is hard
to solve these equations in general, and the averaging procedure faces
analytical difficulties. However, if we make truncated series approximations
for the underlying functions for the case of $\epsilon=0$, we can
in principle carry out averaging. Although the calculations are tedious,
in this paper we show that a remarkably simple 2D averaged phase portrait
is obtained independent of the forcing parameter $q$. And by this
route, a universal averaged dynamics of the Fokker-Planck equation
for small damping and diffusion coefficient is obtained. This universal
form is sufficiently transparent that a complete characterization
of all solutions, for all $q$, is obtained. In particular, in the
2D phase portrait, we have two regions of initial conditions. From
one region, all solutions are unbounded. From the other region, all
solutions go to a unique stable fixed point. The boundary between
these two regions is a separatrix, in the shape of a parabola.

For $\epsilon=0$, Eq. \eqref{eq:ABC-Scaled-Eqn} loosely resembles
the parametrically forced linear Mathieu's differential equation.
It can be shown (e.g., through numerically evaluated Floquet multipliers)
that the $\epsilon=0$ system is stable for $0<q<1.816$ (approximately).
We first obtain the solutions for the $\epsilon=0$ system, and use
them with the method of averaging to study the solutions for $\epsilon>0$.
We convert the system equations into the Lagrange standard form, and
carry out time-averaging of the order $\epsilon$ terms to obtain
the autonomous slow-flow equations. Later, we study the slow-flow
equations along with direct numerical solutions of the original system.
Instead of simultaneously studying variables $A$, $B$, and $C$,
we combine the three equations into a single equation, eliminating
$B$ and $C$, and study the resulting single equation,
\begin{align}
\dddot{A}+4f\dot{A}+2\dot{f}A\qquad\qquad\qquad\qquad\qquad\qquad\label{eq5}\\
+\epsilon\left(12\ddot{A}A+14\dot{A}^{2}-3\ddot{A}-4fA+8fA^{2}\right)+{\cal O}\left(\epsilon^{2}\right) & =0.\nonumber 
\end{align}
Henceforth, we study only Eq.\ \eqref{eq5} and use the method of
averaging, asymptotically valid for sufficiently small $\epsilon$.
To begin with, we present the solutions of Eq.\ \eqref{eq5} for
the collisionless case, $\epsilon=0$.

\section{Solution of the collisionless system, $\mathbf{\epsilon=0}$\label{sec:Solution-of-the-Collisionless}}

The $\epsilon=0$ system is 
\begin{equation}
\dddot{A}+4f\dot{A}+2\dot{f}A=0.\label{eq6}
\end{equation}
The divergence of Eq. \eqref{eq:ABC-Scaled-Eqn} is $3\epsilon(1-4A)$,
which is zero for $\epsilon=0$. This means that the product of the
Floquet multipliers of the unperturbed system must be unity. One of
the multipliers is $+1$ and the other two are complex conjugates,
each of unit magnitude. The unit Floquet multiplier indicates that
there exists a solution that has the same periodicity ($\pi$) as
the forcing function ($f\left(t\right)=-q\cos(2t)$, taking $\omega=2$).
The other two solutions can be aperiodic (when the complex Floquet
multipliers' arguments are irrational fractions of $\pi$) \cite{Shah2008,Banerjee}.
Hence, we need to construct three independent solutions, $\phi_{i}$,
and the general solution of Eq.\ \eqref{eq6} can be written as 
\begin{equation}
A=a_{1}\phi_{1}+a_{2}\phi_{2}+a_{3}\phi_{3},\label{Aa}
\end{equation}
where $a_{i}$ are arbitrary constants. It can be shown that the $\phi_{i}$
can be written using a trigonometric series \cite{Shah2008} (also
see \cite{Abraham,McLachlan}), 
\begin{eqnarray}
\phi_{1} & = & \sum_{k=-N}^{k=N}B_{k}\cos(2kt),\label{phi_1}\\
\phi_{2} & = & \sum_{m=-N}^{m=N}C_{m}\cos\left(\left(\beta+2m\right)t\right),\label{phi_2}\\
\phi_{3} & = & \sum_{n=-N}^{n=N}D_{n}\sin\left(\left(\beta+2n\right)t\right).\label{phi_3}
\end{eqnarray}
where the coefficients $B_{k}$, $C_{m}$, $D_{n}$ and the frequency,
$\beta$, are functions of $q$ and $N$. $N=\infty$ is needed for
an exact solution, but convergence is rapid and a moderate value of
$N$ is good enough for practical purposes \cite{McLachlan}. 

To determine $\beta$, we do the following. We first substitute $\phi_{2}$
given by Eq.\ \eqref{phi_2} in Eq.\ \eqref{eq6}. The resulting
expressions are then simplified, and the coefficient of each harmonic,
i.e. coefficients of $\cos\left(\left(\beta+2m\right)t\right)$ with
$m=-N\ldots N$, are set to zero. The obtained $2N+1$ equations are
written in matrix form, $Mx=0$, where $M_{ij}$ denotes the coefficient
of $C_{j}$ in the $i^{{\rm th}}$ equation, $i$ and $j$ vary from
$-N$ to $N$, and $x$ here denotes the unknowns $C_{m}$. For a
non-trivial solution the determinant of $M$, which depends on $q$
and the as yet unknown $\beta$, must be zero. For known $q$ we can
solve numerically for $\beta$. For that pair of $q$ and $\beta$,
$M$ has a zero eigenvalue; and the corresponding eigenvector gives
the coefficients $C_{m}$, up to a scalar multiple. We choose $C_{0}=1$
for definiteness. We repeat the procedure for $\phi_{3}$. The equation
for determining $\beta$ for a given $q$ is identical (as it needs
to be; because these continuous functions determine the Floquet multipliers,
which need to be complex conjugates).

\begin{figure}
\centering{}\includegraphics[width=3in]{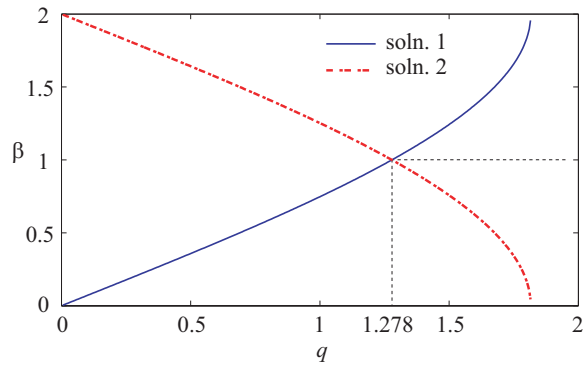} \caption{\label{beta_q_sol} The two solutions obtained for $0<\beta<2$ for
different values of $q\in\left(0,1.816\right)$. We use the solid
curve in our analysis.}
\end{figure}

Proceeding as described above, we vary $q$ between 0 to 1.816. For
each $q$, we obtain multiple roots. The non-uniqueness in $\beta$
can be understood by examining Eqs.\ \eqref{phi_1} through \eqref{phi_3},
with $N=\infty$. In such a case, clearly any $\beta$ can be replaced
by $\pm\beta+2k$ for any integer $k$. For this reason, we restrict
attention to $0<\beta<2$. Even this restricted case has two solutions
for any $q$. These two solutions for different $q$ are shown in
Fig.\ \ref{beta_q_sol}. The solution curves cross at $q\approx1.278$,
when $\beta=1$. We use the $\beta$ corresponding to the solid line
in the figure for definiteness. It turns out that the coefficients
$C_{m}=D_{n}$. Finally, $\phi_{1}$ is determined by inserting $\beta=0$
for all $q$. It is found that the sines give an identically zero
solution; and the cosines give the third and final linearly independent
nontrivial solution. The coefficients $B_{k}$ are found without ambiguity,
again with $B_{0}=1$. In the next section, we consider the collisional
case, $0<\epsilon\ll1$, and analyze the equations using the method
of averaging.

\section{Averaging of the collisional equations, $\mathbf{0<\epsilon\ll1}$\label{sec:Averaging-of-the-Collisional}}

We adopt the method of averaging to study Eq.\ \eqref{eq5} for $0<\epsilon\ll1$.
In order to carry out averaging, the equations need to be represented
in the Lagrange standard form \cite{Verhulst}, which is 
\begin{equation}
\dot{h}=\epsilon g({h},t)+{\cal O}\left(\epsilon^{2}\right).\label{lst}
\end{equation}
In the above, $h$ is a vector of unknowns, and the $t$-dependence
of $g$ is typically oscillatory although not necessarily perfectly
periodic. Note that the right-hand side is ${\cal O}\left(\epsilon\right)$.
The slow flow equations are 
\[
\dot{h}_{{\rm av}}=\epsilon\bar{g}
\]
where 
\[
\bar{g}=\lim_{T\to\infty}\frac{1}{T}\int_{0}^{T}g({h_{{\rm av}}},t)dt,
\]
where in the right hand side, $h_{{\rm av}}$ is treated as a constant
while integrating with respect to time. The averaged dynamics of $h_{{\rm av}}$
approximates the original $h$ over time scales of ${\cal O}\left(1\big/\epsilon\right)$.
The averaging can be done over one period only, if $g$ is periodic
in $t$; and over several discrete periods, term-wise, if $g$ is
the sum of individual periodic terms with incommensurate periods;
and over a long time with a limit, as indicated, in the general case.
To solve Eq.\ \eqref{eq5}, we drop terms of $\mathcal{O}\left(\epsilon^{2}\right)$
and higher and begin with Eq.\ \eqref{Aa}, but with the $a_{i}$
now being treated as time-varying coordinates,
\begin{equation}
A\left(t\right)=a_{1}\left(t\right)\phi_{1}\left(t\right)+a_{2}\left(t\right)\phi_{2}\left(t\right)+a_{3}\left(t\right)\phi_{3}\left(t\right).\label{eq:newsolform}
\end{equation}

Differentiating the above once with respect to time, we obtain 
\begin{equation}
\dot{A}=\dot{a}_{1}\phi_{1}+\dot{a}_{2}\phi_{2}+\dot{a}_{3}\phi_{3}+a_{1}\dot{\phi}_{1}+a_{2}\dot{\phi}_{2}+a_{3}\dot{\phi}_{3}.
\end{equation}
Here, noting that we have three $a_{i}$ in place of one original
unknown $A$, we must adopt two constraint equations. The choice of
these constraint equations is classical and routine, and is given
by

\begin{align}
\dot{a}_{1}\phi_{1}+\dot{a}_{2}\phi_{2}+\dot{a}_{3}\phi_{3} & =0\nonumber \\
\dot{a}_{1}\dot{\phi}_{1}+\dot{a}_{2}\dot{\phi}_{2}+\dot{a}_{3}\dot{\phi}_{3} & =0\label{eq:Contraints}
\end{align}
which results in the following equations for the first and second
derivatives of $A$,
\begin{align*}
\dot{A} & =a_{1}\dot{\phi}_{1}+a_{2}\dot{\phi}_{2}+a_{3}\dot{\phi}_{3}\\
\ddot{A} & =a_{1}\ddot{\phi}_{1}+a_{2}\ddot{\phi}_{2}+a_{3}\ddot{\phi}_{3}
\end{align*}
Note that since $\phi_{i}$ is a solution of the $\epsilon=0$ system,
it satisfies 
\[
a_{i}\left(\dddot{\phi}_{i}+4f\dot{\phi}_{i}+2\dot{f}\phi_{i}\right)=0,
\]
identically for each $i$, and for any value of $a_{i}$, irrespective
of whether $a_{i}$ is time varying or not.

We next substitute Eq.\ \eqref{eq:newsolform} in Eq.\ \eqref{eq5},
and use Eq.\ \eqref{eq:Contraints}, to obtain 

\begin{equation}
\dot{a}_{1}\ddot{\phi}_{1}+\dot{a}_{2}\ddot{\phi}_{2}+\dot{a}_{3}\ddot{\phi}_{3}=-\epsilon w\label{meq3}
\end{equation}
where 
\begin{eqnarray}
w & = & -\Big\{12\left(a_{1}\ddot{\phi}_{1}+a_{2}\ddot{\phi}_{2}+a_{3}\ddot{\phi}_{3}\right)\left(a_{1}\phi_{1}+a_{2}\phi_{2}+a_{3}\phi_{3}\right)\nonumber \\
 &  & +14\dot{f}\left(a_{1}\dot{\phi}_{1}+a_{2}\dot{\phi}_{2}+a_{3}\dot{\phi}_{3}\right)^{2}\nonumber \\
 &  & -3\left(a_{1}\ddot{\phi}_{1}+a_{2}\ddot{\phi}_{2}+a_{3}\ddot{\phi}_{3}\right)-4f\left(a_{1}\phi_{1}+a_{2}\phi_{2}+a_{3}\right)\nonumber \\
 &  & +8f\left(a_{1}\phi_{1}+a_{2}\phi_{2}+a_{3}\phi_{3}\right)^{2}\phi_{3}\Big\}.
\end{eqnarray}
Combining Eqs.\  \eqref{eq:Contraints} and \eqref{meq3}, we obtain
\begin{equation}
\begin{bmatrix}\phi_{1} & \phi_{2} & \phi_{3}\\
\dot{\phi}_{1} & \dot{\phi}_{2} & \dot{\phi}_{3}\\
\ddot{\phi}_{1} & \ddot{\phi}_{2} & \ddot{\phi}_{3}
\end{bmatrix}\begin{Bmatrix}\dot{a}_{1}\\
\dot{a}_{2}\\
\dot{a}_{3}
\end{Bmatrix}=\begin{Bmatrix}0\\
0\\
-\epsilon w
\end{Bmatrix},\label{tempeq1}
\end{equation}
which can be written as 
\begin{equation}
\begin{Bmatrix}\dot{a}_{1}\\
\dot{a}_{2}\\
\dot{a}_{3}
\end{Bmatrix}=-\frac{\epsilon}{\Delta}\begin{Bmatrix}\left(\phi_{2}\dot{\phi}_{3}-\phi_{3}\dot{\phi}_{2}\right)w\\
\left(\phi_{3}\dot{\phi}_{1}-\phi_{1}\dot{\phi}_{3}\right)w\\
\left(\phi_{1}\dot{\phi}_{2}-\phi_{2}\dot{\phi}_{1}\right)w
\end{Bmatrix},\label{pre_avg}
\end{equation}
where $\Delta$ denotes the determinant of the coefficient matrix
on the left hand side of Eq.\ \eqref{tempeq1}. Conveniently, although
this determinant depends on $q$, it is time-invariant for this problem,
and can be calculated at $t=0$ using the computed $\phi_{i}(t)$
and their derivatives. Equation \eqref{pre_avg} is in the Lagrange
standard form, and can be averaged. Fortunately and interestingly,
some big simplifications are possible.

\begin{figure*}[!h]
\centering{}\includegraphics[width=5in]{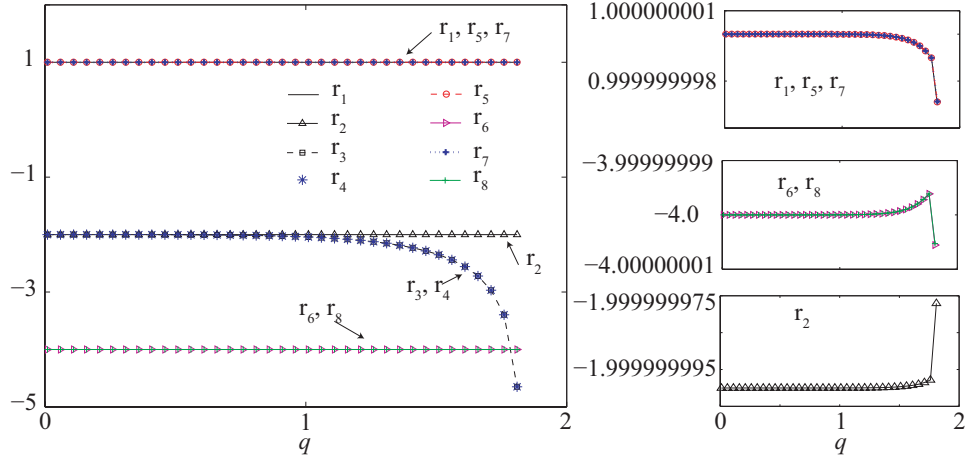} \caption{\label{coeff3} The numerically evaluated coefficients $r_{i}(q)$
of Eq. \eqref{eq:Slow-Flow-Equations} fo various values of $q\in\left(0,1.816\right)$.
Here, we have used $N=8$. As can be clearly seen in the graph, $r_{1},r_{5},r_{7}\approx1$,
$r_{2}\approx-2$, $r_{6},r_{8}\approx-4$ and $r_{3}=r_{4}\in\left(-5,-2\right)$.
This finally gives the simplified slow-flow equations given by Eq.
\eqref{eq:Slow-Flow-Equations1}. The subplot on the right support
our conclusion that, except for $r_{3}$ and $r_{4}$, the other coefficients
are in fact integers and larger $N$ lowers the deviation from integer
values. }
\end{figure*}

\begin{figure*}[!h]
\centering{}\includegraphics[width=5in]{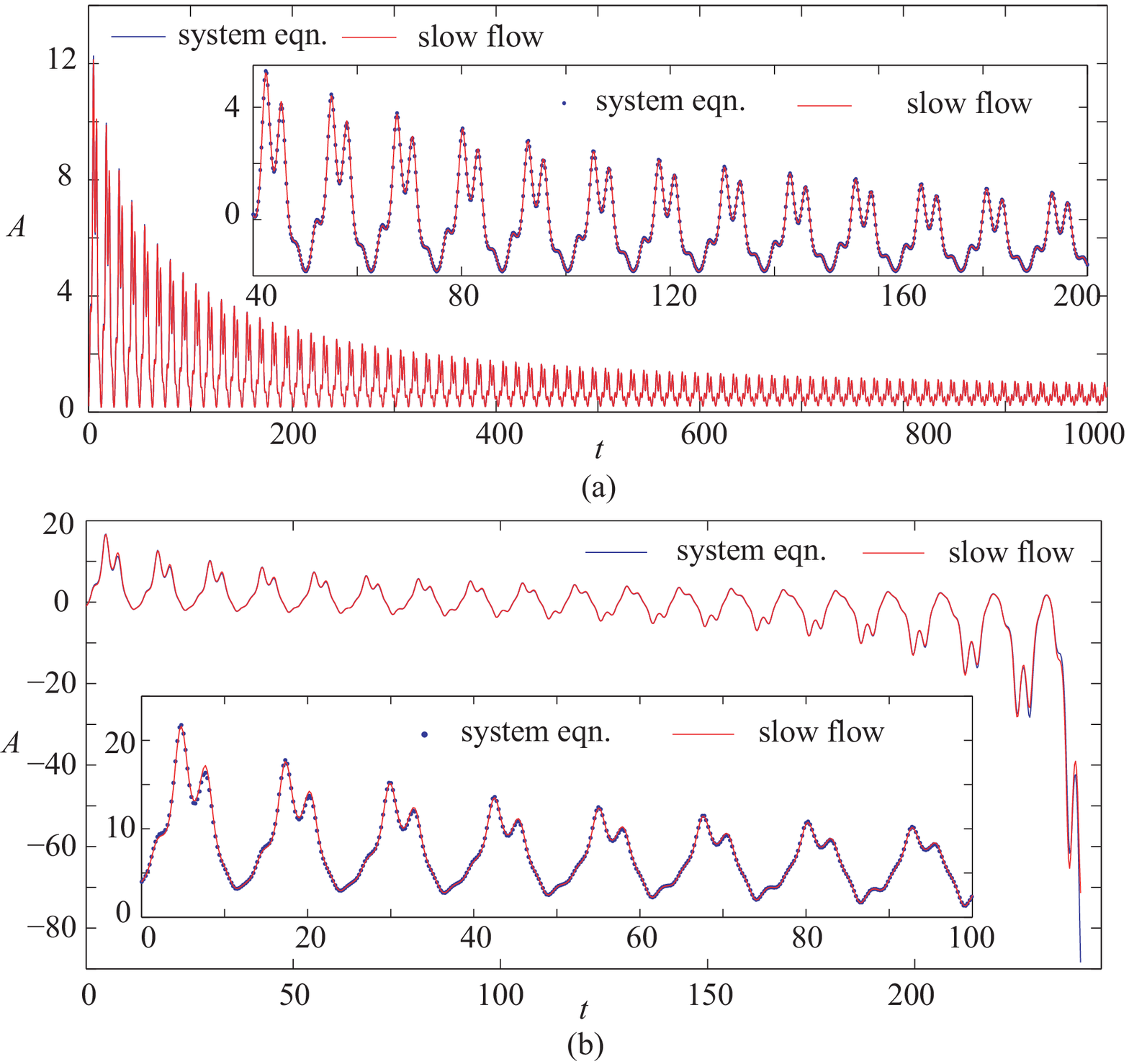} \\
 \caption{\label{val1} Comparison of the solutions obtained by direct numerical
integration of Eq.\ \eqref{eq5} and numerical integration of the
slow-flow equations, Eq. \eqref{eq:Slow-Flow-Equations1}. Parameters
used: $\beta=1\big/2$, $q=0.6899$ and $\epsilon=0.001$. The upper
graph (a) depicts an asymptotically stable trajectory and the lower
graph (b) an unstable one.}
\end{figure*}

\begin{figure*}[!]
\centering{}\includegraphics[width=5in]{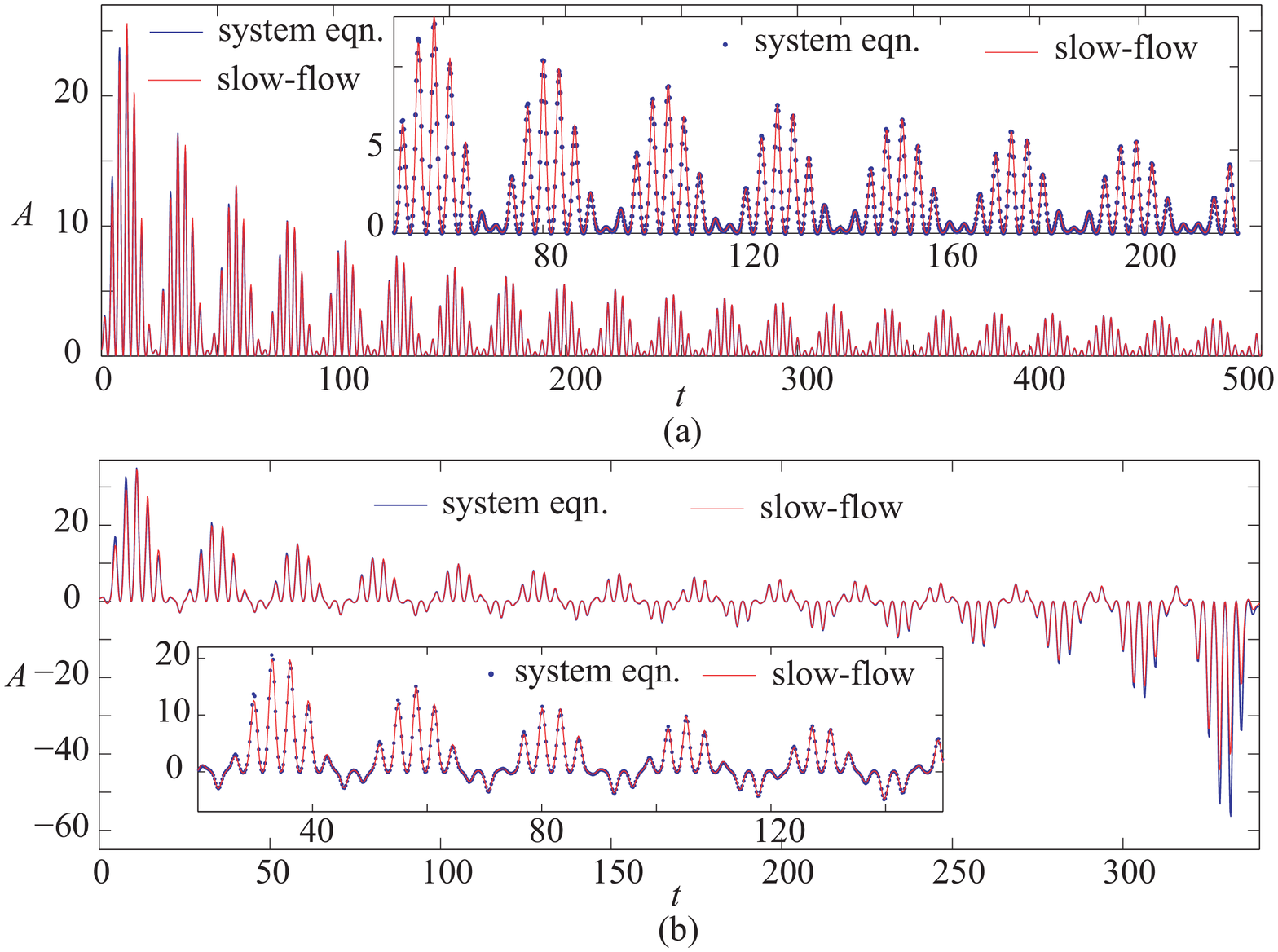} \caption{\label{val2} Comparison of the solutions obtained by direct numerical
integration of Eq.\ \eqref{eq5} and numerical integration of the
slow-flow equations, Eq. \eqref{eq:Slow-Flow-Equations1}. Parameters
used: $\beta=\sqrt{3}$, $q=1.775$ and $\epsilon=0.001$. The upper
graph (a) depicts an asymptotically stable trajectory and the lower
graph (b) an unstable one.}
\end{figure*}

There are many terms in Eq. \eqref{pre_avg}, and each of those terms
is a product of a few functions, and each function is expanded in
a series. However, using various symmetry arguments and trigonometric
identities and simplifications, we can theoretically show that many
of the terms above must have average values equal to zero. This is
a tedious step, and details are omitted. After such simplification,
a fairly large number of terms still remain, and they can be averaged
either numerically, or using symbolic algebra (MAPLE), with numerical
values of the coefficients $B_{k}$, $C_{m}$ and $D_{n}$ along with
$\beta$, determined separately for each $q$.

After such simplification and averaging, including therein division
by $\Delta$, which is constant with respect to time but depends on
$q$, the remaining non-zero averages yield a final slow flow of the
form 
\begin{eqnarray}
\dot{a}_{1} & = & \epsilon\left\{ r_{1}(q)\,a_{{1}}+r_{2}(q)\,{a_{{1}}}^{2}+r_{3}(q)\,{a_{{2}}}^{2}+r_{4}(q)\,{a_{{3}}}^{2}\right\} ,\nonumber \\
\dot{a}_{2} & = & \epsilon\left\{ r_{5}(q)\,a_{{2}}+r_{6}(q)\,a_{{1}}a_{{2}}\right\} ,\nonumber \\
\dot{a}_{3} & = & \epsilon\left\{ r_{7}(q)\,a_{{3}}+r_{8}(q)\,a_{{1}}a_{{3}}\right\} ,\label{eq:Slow-Flow-Equations}
\end{eqnarray}
where we see that several possible quadratic terms have dropped out.
Interestingly, most of the $r_{i}(q)$ in the above equation apparently
have integer values if $N$ is large enough, independent of $q$ (or
$\beta$)! The integer nature of these coefficients is hidden by the
complex expressions involved, but the numerical evidence seems beyond
the realm of coincidence. In Fig. \ref{coeff3}, several of the coefficients
are seen to be extremely close to integer values. We henceforth take
these averaged coefficients as integers. There remains only \emph{one}
non-integer averaged coefficient, henceforth denoted by $\mu(q)$,
which takes values between $2$ and $5$ approximately, and which
appears twice in the slow flow, as $r_{3}(q)$ and $r_{4}(q)$. The
final slow flow has a simple form: 
\begin{eqnarray}
\dot{a}_{1} & = & \epsilon\left(a_{{1}}-2{a_{{1}}}^{2}-\mu(q)\,{a_{{3}}}^{2}-\mu(q)\,{a_{{2}}}^{2}\right)\nonumber \\
\dot{a}_{2} & = & \epsilon\left(a_{{2}}-4a_{{1}}a_{{2}}\right)\nonumber \\
\dot{a}_{3} & = & \epsilon\left(a_{{3}}-4a_{{1}}a_{{3}}\right).\label{eq:Slow-Flow-Equations1}
\end{eqnarray}

We now validate the averaging calculation by comparing two solutions:
(i) direct numerical integration of the original, Eq.\ \eqref{eq5};
and (ii) numerical integration of the slow-flow equations given by
Eq. \eqref{eq:Slow-Flow-Equations1}, followed by inserting the $a_{i}$
in Eq.\ \eqref{eq:newsolform}. Figure\ \ref{val1} shows the solutions
thus obtained for $q=0.6899$, $\beta=1\big/2$, $\epsilon=0.001$,
$N=5$, and $\mu(0.6899)=2.008$. The upper subplot shows a bounded
solution; and the lower subplot, with different initial conditions,
shows an unbounded solution. Both solutions are actually plotted using
two curves, one from the original equations and one from the slow
flow. A zoomed view of a portion is also shown to display the excellent
match between the two solutions, for small $\epsilon$. Figure \ref{val2}
shows a similar match for $\beta=\sqrt{3}$ with $q=1.775$, $\epsilon=0.001$,
$N=5$, and $\mu(1.775)=3.595$. The match, again, is excellent. Note
that the waveforms are different in Figs. \ref{val1} and \ref{val2}
because of differences in the $\phi_{i}$, although the slow flow
in terms of coordinates $a_{i}$ is almost the same. Having established
that the slow flow equations given by Eq. \eqref{eq:Slow-Flow-Equations1}
do indeed capture the dynamics of the original system very well, we
now proceed to study the slow flow equations themselves in the next
section. They allow further simplification.

\section{Dynamics of the slow flow equations\label{sec:Dynamics-of-the-Slow-Flow}}

Writing $\tau=\epsilon t$ (a slow time), we can remove $\epsilon$
from explicit consideration. We multiply the equation for $\dot{a}_{2}$
in Eq. \eqref{eq:Slow-Flow-Equations1} by $a_{2}$ and that for $\dot{a}_{3}$
by $a_{3}$, and add them to obtain 
\begin{eqnarray}
\frac{d}{d\tau}\left(\frac{a_{2}^{2}}{2}+\frac{a_{3}^{2}}{2}\right) & = & \left(a_{2}^{2}+a_{3}^{2}\right)\left(1-4a_{1}\right).
\end{eqnarray}
Writing $Z=\mu\left(q\right)\left(a_{2}^{2}+a_{3}^{2}\right)$, where
$Z\geq0$, we find that Eq. \eqref{eq:Slow-Flow-Equations1} collapse
to a two dimensional parameter-free form, 
\begin{eqnarray}
\dot{a}_{1} & = & -Z+a_{1}\left(1-2a_{1}\right),\label{sfr3}\\
\dot{Z} & = & 2Z\left(1-4a_{1}\right),\label{sfr4}
\end{eqnarray}
where the overdots denote derivatives with respect to $\tau=\epsilon t$.
Eqs.\ \eqref{sfr3} and \eqref{sfr4} have a single, simple, averaged
form that characterizes the dynamics of the original Eq.\ \eqref{eq:ABC-Scaled-Eqn}
for small $\epsilon$. The actual value of $\mu$, and hence $q$,
has no effect on the qualitative dynamics. The averaged slow flow
equation, after this simple transformation, is the same for all $q$
values.

\begin{figure*}
\centering{}\includegraphics[width=5in]{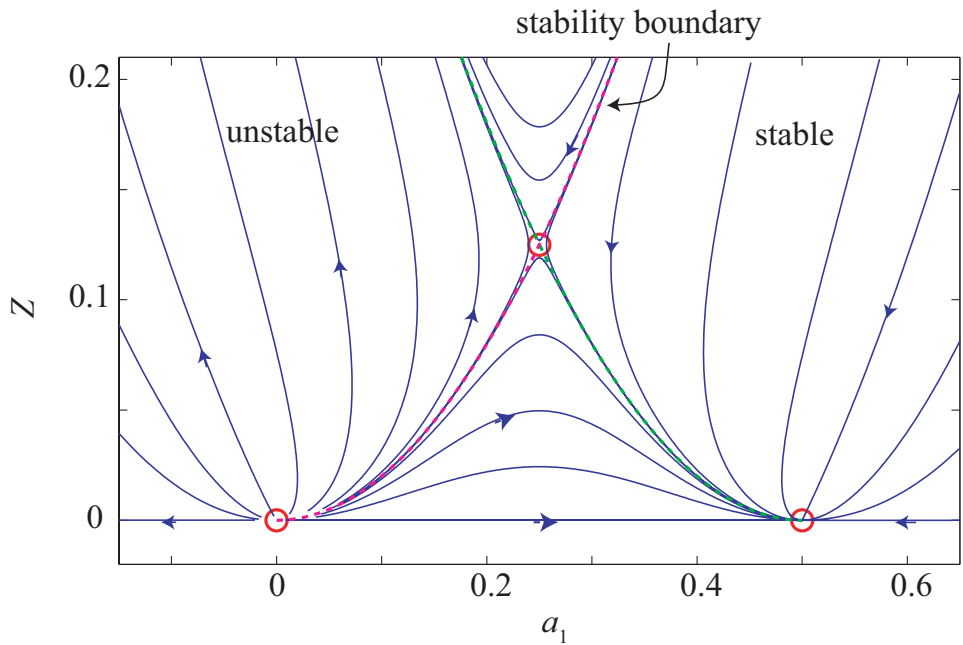} \caption{\label{sfr} Unified phase portrait of Eqs.\ \eqref{sfr3} and \eqref{sfr4}
in the reduced $a_{1}$-$Z$ plane with $Z\geq0$. The fixed points
(red circles) are: (i) (0,0), unstable, (0.5,0) stable, and (iii)
(0.25,0.125), saddle. The dashed curve in magenta, $Z=2a_{1}^{2}$,
denotes the stability boundary. }
\end{figure*}

Fixed points of Eqs.\ \eqref{sfr3} and \eqref{sfr4} are easy to
find. There are three of them: (i) $\left(a_{1},Z\right)=(0,0)$,
unstable node, (ii) ${\displaystyle \left(0.5,0\right)}$, stable
node, and (iii) ${\displaystyle \left(0.25,0.125\right)}$, saddle.
Figure \ref{sfr} shows the phase portrait for Eqs.\ \eqref{sfr3}
and \eqref{sfr4} and for $Z\geq0$. The circles therein denote the
fixed points. The blue lines denote numerically integrated solutions.
The dashed lines denote two invariant manifolds. These invariant manifolds
are parabolas. To obtain one of them, assume $Z=ka_{1}^{2}$. For
this parabola to be an invariant manifold, all points on it must have
$\dot{Z}$ and $\dot{a}_{1}$ in the proportion $2ka_{1}$. It is
easy to check that this condition is satisfied for $k=2$. The other
invariant manifold, a shifted parabola, is similarly found. In the
phase portrait of Fig.\ \ref{sfr}, the dashed curve in magenta,
$Z=2a_{1}^{2}$, is the stability boundary. Solutions that originate
on right of the boundary are stable, while those on the left are unbounded.
All stable solutions settle on $a_{1}=0.5$, and $Z=0$ or $a_{2}=a_{3}=0$.
As $A=\sum_{i}a_{i}\phi_{i}$ (recall Eq.\ \eqref{eq:newsolform}),
this means that solutions of the original system, Eq.\ \eqref{eq:ABC-Scaled-Eqn},
either \emph{grow} without bound, or \emph{settle} to $\pi$-periodic
oscillations (corresponding to a purely $\phi_{1}$ solution). This
conclusion holds for all $0<q<1.816$.

\section{Discussion and conclusions\label{sec:Discussion-and-conclusion}}

In this paper, we have analyzed the Fokker-Planck equation for a plasma
in a Paul trap and found that the averaged equations satisfy the dynamical
behavior shown in Fig. \ref{sfr} irrespective of the value of the
parameter $q\in\left(0,1.816\right)$. This is a significant result
since it is usually possible to analyze the plasma distribution function
only for small values of the system parameters \cite{Dutta,Shah2008}.
Out of all the possible averaged solutions corresponding to different
initial values of $a_{1}$ and $Z$ in Fig. \ref{sfr}, determining
which ones are physically meaningful requires considerable further
analysis since the functions $A,B$ and $C$ have to be analyzed for
all $q$, different initial conditions in a three-dimensional space
and as a function of time. However, the important point to note here
is that, in any case, the form of the solution we have chosen in Eq.
\eqref{eq:FP-Sol-Form1} admits only one unique stable time-periodic
solution. Hence, irrespective of their initial values and/or parameter
$q$, the actual time-varying functions $A,B$ and $C$ can asymptotically
either reach the unique stable time-periodic solution or diverge to
infinity. This result holds true for all values of $q\in\left(0,1.816\right)$.
We have checked that for the stable time-periodic solution at $a_{1}=0.5$
and $Z=0$ as shown in Fig. \ref{sfr}, the corresponding initial
conditions satisfy $A>0$, $C>0$ and $4AC-B^{2}>0$, which shows
the existence of a normalizable initial distribution corresponding
to this solution. This implies that the initial distribution in a
small neighborhood around this time-periodic solution will also be
normalizable and asymptotically reach this same stable solution. 

One may ask what happens to the solutions if we chose a form different
from that given in Eq. \eqref{eq:FP-Sol-Form1}. The question is not
easy to answer, because any such assumed form leads to new equations,
and new constraints which must be checked. The area is certainly promising
for future research. However, for the case of a spatially linear electric
field, the resulting solutions are likely to be qualitatively similar
to that obtained in this paper. Namely, a set of initial conditions
will perhaps converge to the same stable time-periodic solution given
in this paper and another set of initial conditions will lead to divergent
solutions. 

A more important question is, what happens if the trapping field itself
is nonlinear? In such cases, the entire approach will require rethinking.
The lack of an underlying linear system will make superposition impossible;
the possibility of amplitude-dependent solution frequencies will make
resonances more complex and numerous; and the dynamics of even a single
ion will involve significant challenges. In such a situation, the
dynamics of a cloud of ions may well exhibit new and complex phenomena
quite different from the remarkably simple unifying dynamics that
has been discovered in this paper \cite{ShahPhD}.
\begin{acknowledgments}
KS would like to thank Swadhin Agrawal and Ritesh Singh for doing
some preliminary work on this problem. 
\end{acknowledgments}

\end{document}